\documentclass[twocolumn,runningheads]{svjour2}
\smartqed  
\usepackage{graphicx}
\journalname{Astrophysics and Space Science}
\begin{document}

\title{GLAST Large Area Telescope Multiwavelength Planning}



\author{O. Reimer \and 
		P.~F. Michelson \and  
		R.~A. Cameron \and
        S.~W. Digel \and \\
        D.~J. Thompson \and
        K.~S. Wood  }

\authorrunning{O. Reimer et al.} 

\institute{O. Reimer and P.~F. Michelson\at
              Stanford University, HEPL \& KIPAC \\
              Stanford, CA 94305, USA \\
              \email{olr@stanford.edu}           
           \and
              R.~A. Cameron and S.~W. Digel\at
              Stanford University, SLAC \& KIPAC \\
              Menlo Park, CA 94025, USA \\
           \and
              D.~J. Thompson \at
              NASA Goddard Space Flight Center \\
              Greenbelt, MD 20771, USA \\
            \and
		  	  K.~S. Wood \at
              Naval Research Laboratory \\
              Washington, DC 20375, USA \\
              }

\date{Received: date / Accepted: date}

\maketitle

\begin{abstract}
Gamma-ray astrophysics depends in many ways on multiwavelength studies. The 
Gamma-ray Large Area Space Telescope (GLAST) Large Area Telescope (LAT) 
Collaboration has started multiwavelength planning well before the scheduled 
2007 launch of the observatory. Some of the high-priority multiwavelength needs 
include: (1) availability of contemporaneous radio and X-ray timing of pulsars; 
(2) expansion of blazar catalogs, including redshift measurements; (3) improved 
observations of molecular clouds, especially at high galactic latitudes; 
(4) simultaneous broad-band blazar monitoring; (5) characterization of 
gamma-ray transients, including gamma ray bursts; (6) radio, optical, X-ray and 
TeV counterpart searches for reliable and effective sources identification and 
characterization. Several of these activities are needed to be in place before 
launch. 
\end{abstract}

\section{Introduction}
\label{intro}
Multiwavelegth (MW) observations are important for the GLAST observatory \cite{PM03}.  
Particular motivations for such studies with the LAT include: 

\begin{itemize}

\item Source identification and population studies.

\item Intensive exploration of the brightest and most variable sources that will allow deep study of the source physics.

\item Rapid follow-up on transients (e.g. gamma-ray bursts (GRBs), blazar flares). The GLAST mission is designed to support rapid notification for follow-up.

\item Understanding the high-energy diffuse emission of the Milky Way

\end{itemize}
The GLAST LAT Multiwavelength Coordination Group (GLAMCOG) has been formed to prioritize 
science-driven needs and develop an implementation plan for cooperative multiwavelength 
observations before and during the GLAST mission. This work is being coordinated with the 
GLAST Burst Monitor (GBM) and GLAST Project science teams. 
Some of the known multiwavelength needs are described here, along with the steps being taken 
to meet those needs. This work is preliminary and does not represent the full range of 
multiwavelength activities that will be investigated. 

\section{Science Goals Needing Multiwavelength Studies}
\label{sec:1}
Based on current knowledge of the gamma-ray sky, many GLAST LAT studies will require MW 
cooperation for the maximum scientific return.  Four of these are described here. Table 1 
summarizes the ways GLAST will use MW observations to address these issues.

\subsection{Probing the Extragalactic Background Light (EBL) with Blazars}

The EBL contains unique information about the epochs of formation and the evolution of 
galaxies and in what environments the stars of the universe formed. Direct EBL measurements, 
however, require accurate model-based subtraction of bright foregrounds (e.g., zodiacal light). 
An alternative approach is to extract the imprint of EBL absorption, as a function of redshift, 
from high-energy spectra of extragalactic sources.  GeV Gamma rays seen by the LAT interact with 
the optical EBL.  Blazars, a known gamma-ray source class, provide the source photons.  
This approach is only feasible on a statistical basis and therefore requires detailed broad-band 
spectral measurements and redshifts for a large sample of blazars, probably more than the known 
blazar population\cite{AC04}.   
      
\subsection{Modeling the Diffuse Gamma-Ray Emission from the Milky Way}

The diffuse Galactic gamma-ray emission must be well characterized for analysis of LAT data, 
much more so than for EGRET, owing to the LAT's vastly better statistics and better angular 
resolution. The origin of this radiation is predominantly cosmic-ray interactions with 
interstellar gas and the interstellar radiation field.  Information about these cosmic-ray 
targets can only come from multiwavelength observations and analysis (e.g. \cite{AS00}. 
Some fundamental questions remain from EGRET, with results limited by knowledge of the diffuse emission; e.g.
\begin{itemize}

\item What is the nature of the source in the Galactic Center region?

\item  What is the origin(s) of the isotropic gamma-ray background?

\item  Is there gamma-ray evidence for particle dark matter?

\end{itemize}

\subsection{Identifying New Source Classes}

About half the sources in the third EGRET catalog \cite{RH99} remain unidentified, largely because 
the error boxes were too large for identifying a unique counterpart in deep searches at other 
wavelengths. Potential new source classes include starburst galaxies, radio galaxies, clusters of galaxies, 
pulsar wind nebulae, colliding winds of massive stars, and microquasars.  The major increase in sensitivity 
and better angular resolution of GLAST LAT (especially at higher energies) will produce much smaller error boxes, 
sub-arcmin in many cases. Finding new source classes is an important part of the discovery potential of the LAT.

\subsection{Investigating Physics of Extreme Conditions in Pulsars}

Pulsars, rotating neutron stars, are sites of particle acceleration and interactions in 
extreme gravitational, electric, and magnetic fields.  Multiwavelength pulsar studies have 
shown significant variety of pulsar properties as a function of wavelength\cite{DT04}. 
A key to using gamma rays to help decipher these extreme processes is having accurate, 
absolute timing data for many pulsars.
With the exception of a few X-ray pulsars, the radio band provides the timing information 
needed by observations across the spectrum. A sizeable radio timing program is beyond the 
scope of routine radio pulsar programs.

\section{Planning for GLAST LAT Multiwavelength Studies}
\label{sec:2}
A variety of approaches for MW research will be used.  Four of these are outlined here.

\subsection{Coordinated Multiwavelength Campaigns}

Particularly for time-variable emission phenomena as, e.g. in blazars, coordinated observations across the 
electromagnetic spectrum provide essential information about locations and processes of particle acceleration 
and interaction. Examples are campaigns by the Whole Earth Blazar Telescope (WEBT) \cite{MB05}. The GLAST LAT 
team will be an active participant in such campaigns. Because LAT will serve as an all-sky monitor, it will be 
an important trigger for coordinated efforts. The GLAST Project is a co-sponsor of the Global Telescope Network 
({\it http://gtn.sonoma.edu/public/}), an informal association of small telescopes supporting such observations.

\subsection{Wider and Deeper Surveys for Molecular Gas}

Despite extensive surveys, our knowledge of the matter and radiation fields in the Milky way remains incomplete.  
Two important sets of observations are needed in order to upgrade the model of the diffuse gamma-ray emission 
in the Galaxy:

\begin{itemize}

\item Extend CO surveys uniformly to high latitudes - CO is a tracer of the important molecular hydrogen component 
of clouds.  Cosmic rays interacting with small molecular clouds will be interpreted as unidentified sources unless 
the matter content of the clouds is modeled properly \cite{DT05}.  Such identified clouds would clearly limit 
dark matter studies.

\item C$^{18}$O observations - This optically thin tracer of molecular hydrogen must be measured in special 
directions (e.g. the Galactic Center and spiral arm tangents) to assess whether velocity crowding is affecting 
calculations of molecular column density, and for carefully pinning down the diffuse emission.

\end{itemize}

\subsection{Sample Strategies for Identifying Gamma-Ray Sources}

Even with the improved error boxes to be provided by LAT compared to EGRET, gamma-ray source identification 
will often involve more than simple positional association.  We note three approaches that can be used to 
assist in this identification process:

\begin{itemize}

\item The ``Top $\rightarrow$ Down'' approach will search LAT error boxes for X-ray counterparts with nonthermal, 
hard spectra, then use the X-ray position to find corresponding optical and radio sources. An example 
of this method is the identification of 3EG~J1835+5918 with the X-ray source RX J1836.2+5925 as a likely 
isolated neutron star \cite{NM01},\cite{OR01}.   A challenge will be proposing for enough X-ray telescope 
time to identify the large number of new sources expected with LAT. 

\item The ``Bottom $\rightarrow$ Up'' approach will search LAT error boxes for radio counterparts with flat spectra, 
then follow up with redshift and polarization measurements in the optical to identify potential blazars. 
An example of this method was the identification of 3EG~J2006$-$2321 as a likely blazar \cite{PW02}. 
The VLBA Imaging and Polarization Survey (VIPS) program is one program studying candidate blazars at 
present\\ ({\it http://www.phys.unm.edu/~gbtaylor/VIPS/}). 

\item Correlated variability between gamma-rays and radio, IR, optical, and/or X-rays will provide one of the 
most distinctive signatures for source identification. Pan-STARRS is one optical facility, well-matched to 
the LAT for correlated studies\\ ({\it http://pan-starrs.ifa.hawaii.edu/public/}). 

\end{itemize}

\subsection{Pulsar Timing and Searches}

Pulsar timing programs at facilities such as Arecibo, Parkes, Jodrell Bank, Nan\c{c}ay, and Green Bank 
are being planned in cooperation with Steve Thorsett, a GLAST Interdisciplinary Scientist. 
After launch, unidentified LAT sources will provide targets for deep radio pulsar searches. Similar 
searches will be needed using X-ray telescopes. 

\begin{table*}[t]
\caption{Summary of Some Multiwavelength Needs and Planning}
\centering
\label{tab:1}       
\begin{tabular}{| p{4cm}| p{4cm}| p{4cm}| p{4cm}|}
\hline\noalign{\smallskip}
{\bf Science Objective} & {\bf What GLAST Provides} & {\bf Multiwavelength Requirements} & {\bf Multiwavelength Planning Activities} \\[6pt]
\tableheadseprule
& & & \\
Differential measurement (vs Z) of extragalactic background light to Z $\sim$5.5 &	Measurement of blazar spectra in band where cutoffs are expected from   $\gamma$ + $\gamma_{gebl}$ $\rightarrow$ e$^+$ + e$^-$ &	Broadband contemporaneous/ simultaneous spectral measurements (radio, optical, X-ray, TeV) of blazar spectra, particularly around the synchrotron peak &
Cooperate with and expand existing multiwavelength blazar and GRB campaigns (e.g. WEBT, ENIGMA, GTN, Swift) to have the broadest possible coverage during the mission \\[6pt]
\cline{1-2}
& & & \\
Resolve origin of particle acceleration and emission mechanisms in systems with relativistic jets, supermassive black holes &	All-sky monitoring coverage of blazar flares and Gamma Ray Bursts (GRB) &	
Radio and optical surveys of flat-spectrum radio sources to extend blazar catalogs, including redshift measurements &	
Participate with and encourage programs to expand blazar catalogs and measure redshifts for flat-spectrum radio sources 
 \\
& & & \\
\hline
& & & \\
Reliable model of Milky Way diffuse emission required for accurate source localization and to facilitate search for dark matter &	Mapping of cosmic ray interactions with all forms of interstellar matter &	Extend CO surveys to high galactic latitude;
&	Promote needed CO and other tracer observations;\\
& & Survey special directions (eg. spiral arms, Galactic Center) with optically thin tracer (e.g.  C$^{18}$O) & Work with observers to reduce data and incorporate into a model of the diffuse gamma-ray emission \\
& & & \\
\hline
& & & \\
Search out and understand new classes of gamma-ray sources &	Large number of source detections; & Counterpart searches at all other wavelengths, and in the Multi-Messenger observation channels; & Identify facilities and plan proposal strategies for obtaining observing time needed to identify gamma-ray sources at other wavelengths; \\
& Relatively uniform sky coverage; & Population studies; &  \\
& &	 &   \\
& Good positions, energy spectra, time histories & Correlated variability; & Cooperate with existing and planned monitoring surveys; \\
& & & \\ 
& & Contemporary, complete astronomical catalogs & Prepare for use of the many available astronomical catalogs and observation facilities \\
& & & \\
& &  Multi-Messenger modeling & \\
& & & \\
\hline
& & & \\
Understand particle acceleration and emission mechanisms in extreme environments of rotating neutron stars &	Spectra and light curves resulting from primary interactions of the most energetic particles &	Contemporaneous radio and X-ray pulsar timing observations &	Select pulsar candidates for radio timing; \\
& & & Work with radio and X-ray astronomers to monitor timing of selected pulsars; \\
& & & \\
& & & Plan proposals for radio and X-ray pulsar searches \\ 
& & & \\
\noalign{\smallskip}\hline
\end{tabular}
\end{table*}

\section{Summary}

The GLAST Large Area Telescope science will be optimized by coordinated multiwavelength observations and analysis.
GLAST welcomes cooperative efforts from observers at all wavelengths.\\ 
See {\it http://glast.gsfc.nasa.gov/science/multi/} for further information. 
To be added to the Gamma-Ray multiwavelength information mailing list, please contact Dave Thompson 
(djt@egret.gsfc.nasa.gov).  The GLAST Guest Investigator program will have opportunities for developmental 
and correlative observations. \\ See {\it http://glast.gsfc.nasa.gov/ssc/proposals/} for further information. 

\acknowledgement{The GLAST Large Area Telescope is an international effort, with U.S. funding provided by the Department of Energy and NASA.}


%
%



\begin{thebibliography}{}
%
%
\bibitem{PM03}  Michelson, P. : Instrument for the Gamma-ray Large Area Space Telescope (GLAST) Mission In: \textit{X-Ray and Gamma-Ray Telescopes and Instruments for Astronomy}, Ed. J. E. Tr\"umper, H. D. Tananbaum (SPIE Vol. 4851, 2003) pp. 1144--1150
  
\bibitem{AC04}
Chen, A., Reyes, L.C., Ritz, S. : Detecting the Attenuation of Blazar Gamma-Ray Emission by Extragalactic Background Light with the Gamma-Ray Large Area Space Telescope. ApJ {\bf 608}, 686--691 (2004)
\bibitem{AS00}
Strong, A. S., Moskalenko, I. V., Reimer, O. : Diffuse Continuum Gamma Rays from the Galaxy. ApJ {\bf 537}, 763--784 (2000)
\bibitem{RH99}
Hartman, R. C. et al. : The Third EGRET Catalog of High-Energy Gamma-Ray Sources. ApJS {\bf 123}, 79--202 (1999)
\bibitem{DT04}
Thompson, D. J.: Gamma Ray Pulsars. In: K.S. Cheng, G. E. Romero (ed.) Cosmic Gamma-Ray Sources. Kluwer Academic Publishers, Dordrecht, The Netherlands,  pp. 149--168 (2004)
\bibitem{MB05}
B\"ottcher, M.  et al. : Coordinated Multiwavelength Observation of 3C 66A during the WEBT Campaign of 2003-2004. ApJ {\bf 631}, 169--186 (2005)
\bibitem{DT05}
Torres, Diego F., Dame, T. M., Digel, S. W. : High-Latitude Molecular Clouds as Gamma-Ray Sources for the Gamma-Ray Large Area Space Telescope. ApJ {\bf 621}, L29--L32 (2005)
\bibitem{NM01}
Mirabal, N., Halpern, J. P.: A Neutron Star Identification for the High-Energy Gamma-Ray Source 3EG J1835+5918 Detected in the ROSAT All-Sky Survey. ApJ {\bf 547}, L137--L140 (2001)
\bibitem{OR01}
Reimer, O. et al. : Multifrequency studies of the enigmatic gamma-ray source 3EG J1835+5918. MNRAS {\bf 324}, 772--780 (2001)
\bibitem{PW02}
Wallace, P. M. et al.: An Active Galactic Nucleus Identification for 3EG J2006-2321. ApJ {\bf 569}, 36--43 (2001)

\end{thebibliography}


\end{document}